\documentclass[sigconf]{aamas}  
\pdfoutput=1

\usepackage{booktabs}
\usepackage{bcprules}
\typicallabel{ClrInt$_4$}
\usepackage{ASL-ASC-JAIR}

\setcopyright{ifaamas}  
\acmDOI{doi}  
\acmISBN{}  
\acmConference[AAMAS'18]{Proc.\@ of the 17th International Conference on Autonomous Agents and Multiagent Systems (AAMAS 2018), M.~Dastani, G.~Sukthankar, E.~Andre, S.~Koenig (eds.)}{July 2018}{Stockholm, Sweden}  
\acmYear{2018}  
\copyrightyear{2018}  
\acmPrice{}  

\newcommand{\FORGET}[1]{}

\renewcommand\footnotetextcopyrightpermission[1]{} 
\begin{document}

\title{Towards a Programmable Framework for Agent Game Playing}




\author{Francis Lawlor}
\affiliation{%
  \institution{UCD School of Computer Science}
  \streetaddress{Belfield}
  \city{Dublin} 
  \state{Ireland} 
}
\email{francis.lawlor@ucdconnect.ie}
\author{Rem Collier}
\affiliation{%
  \institution{UCD School of Computer Science}
  \streetaddress{Belfield}
  \city{Dublin} 
  \state{Ireland} 
}
\email{rem.collier@ucd.ie}
\author{Vivek Nallur}
\affiliation{%
  \institution{UCD School of Computer Science}
  \streetaddress{Belfield}
  \city{Dublin} 
  \state{Ireland} 
}
\email{vivek.nallur@ucd.ie}


\begin{abstract}  
The field of Game Theory provides a useful mechanism for modeling many decision-making scenarios. In participating 
in these scenarios individuals and groups adopt particular strategies, which generally perform with varying levels 
of success.  However, most results have focussed on players that play the same game in an iterated fashion. This 
paper describes a framework which can be used to observe agents when they do not know in advance which game they 
are going to play. That is, the same group of agents could first play a few rounds of the Iterated Prisoner's Dilemma, 
and then a few rounds of the Linear Public Goods Game, and then a few rounds of Minority Game, or perhaps all 
games in a strictly alternating fashion or a randomized instantiation of games. This framework will allow for 
investigation of agents in more complex settings, when there is uncertainty about the future, and limited resources 
to store strategies.
\end{abstract}


\maketitle


\section{\uppercase{Introduction}}
\label{sec:introduction}

\noindent Turocy and von Stengel define Game Theory as ''the formal study of decision-making where several players 
must make choices that potentially affect the interests of the other players''~\cite{Turocy2001}. Game Theory 
provides the ability to reduce real world problems - those which are stylized in the form of a game - to mathematical 
models. This has resulted in the establishment of a huge array of formally recognized problems. Examples include the 
Minority Game~\cite{Challet1997}, the Iterated Prisoner’s Dilemma~\cite{Binmore2006}, Public Goods Game~\cite{Isaac1994}, 
etc. However, most game theoretic approaches consider each game in isolation and do not investigate mechanisms, 
strategies or behaviours across games. This is inadequate for computational social simulations, since in reality 
human beings are confronted by many (possibly conflicting) games, and respond differently to each. Cognitive limitations 
on memory, time, resources, etc. combined with spatial influences result in actual behaviour that is not 
predicted/matched by rational agent simulations~\cite{Thaler2016}. In parallel, the field of AI has been developing 
algorithms and techniques that focus on achieving human-level competence in several fields. Complex games with large 
state spaces such as Checkers~\cite{Schaeffer2007} and Go~\cite{Silver2016}, games with imperfect information such as 
Texas Hold'Em Poker~\cite{Bowling2015} have been successfully played by algorithms. There are even algorithms that 
attempt to induce long-term cooperative behaviour in partly-competitive settings~\cite{Groom2007,Crandall2018}. However, 
to the best of our knowledge, there has been no attempt at finding a general bag of strategies that work in both 
competitive and cooperative settings, allow for the emergence of cooperation, altruism, and are flexible enough to be 
applied to many games. Human beings, on the other hand, have limited cognitive abilities and yet are able to (seemingly) 
effortlessly switch from domain to domain, and successfully compete as well as cooperate.
We believe this is partly due to the inadequacy of computational simulation tools available for complex adaptive 
agents to play multiple games in heterogeneous environments. There are no tools, for example, that allow machine 
learning agents to be in competition with evolutionary agents across multiple heterogeneous games. This paper reports 
on a game-playing framework called \textit{Arena}, that attempts to ameliorate the situation by providing an 
Agent-based reconfigurable environment for tournaments. A tournament, here, is defined to be multiple games played 
in some order, with players encountering each other \textit{across} games. The primary goal of such a reconfigurable 
environment is that researchers can modify players, strategies, game rules, environments independent of each other, 
thus enabling not only a richer simulation but also a controlled increase in complexity. \textit{Arena} aims to 
enable co-evolution of players, allowing the emergence of trust, reputation and coalitions as a natural consequence 
of known player identity across games.

\section{Related Work}
Multi-Agent systems have been used for simulations of many problem domains such as Smart-grids~\cite{Mohsenian-Rad2010}, vehicular ad-hoc networks~\cite{Ding2003}, smart buildings~\cite{Yoon2014}, e-procurement~\cite{Wu2007}, cloud computing~\cite{Nallur2010}, healthcare~\cite{Axisa2005}, and transport~\cite{Masoum2011}. However, many of these domains are complex enough that getting the \textit{same} agent to adapt and perform in a cross-domain manner is very difficult. Desirable properties such as evolutionarily stable equilibria, allocative fairness etc. are also difficult to be formulated across multiple domains in an easily understandable manner. Game theory, on the other hand, has been used to mathematically compute or discover optimizing behaviour across a plethora of games, with a variety of constraints. However, it has not been used evaluate 'realistic' players that often encounter more than one game and suffer from resource constraints and bounded rationality. This paper attempts to create an intersection of game theory and multi-agent systems such that the advances in agent-modelling and simulation techniques can be applied in a well-understood games. Current approaches can roughly be divided into multi-agent simulations and game theory simulations. 

\paragraph{Multi-Agent Simulations:}
Multi-Agent simulation environments are often general purpose environments, that focus on providing ease of modelling of problem domain or agent behaviour or learning strategies. Most simulations tend to involve the creation of a bespoke implementation that is tied closely to the domain it is being evaluated against, with many implementations written in languages such as Netlogo~\cite{Wilensky1999} or environments such as Jason~\cite{Bordini2007}, Jade~\cite{Bellifemine2007}, etc. In spite of the sophistication of the agent environments, there are rarely any principled frameworks to test the \textbf{same} strategy/learning algorithm in \textbf{multiple} scenarios, since modelling multiple problem domains with any fidelity remains onerous. Hence, despite multiple reports of the importance and value of diversity for robustness in multi-agent systems~\cite{Lewis2014,Song2015}, to the best of our knowledge, there are no agent-based simulation platforms that allow for a systematic investigation of the same agent/strategy across heterogeneous problem domains.

\paragraph{Game Theory Simulations:}

Game theory abstracts away from the heterogeneity of problem domains and investigates stylized phenomena where the  variables of interest are: payoffs, player moves and the presence of equilibria in conditions of repeated play.
Most game theory simulators, to the best of our knowledge, focus on certain kinds of games. For instance, Gambit~\cite{McKelvey2016} is an attempt to build a generalized game playing framework for non-cooperative games where all players have access to a common set of strategies and the payoff for various moves is known at design time.  

\section{Generalizing Game Playing}
Current work of game playing tends to focus on the investigation of strategies for playing specific games, finding equilibria in repeated play or proving other properties for a specific game. 
However, games can vary across many dimensions, such as number of players (two-player, multi-player), moves (simultaneous, sequential), payoff (zero-sum, non-zero-sum), duration (repeated, one-shot), etc. In order to build a generalized game playing simulator, it is important to first agree on what constitutes a game and more generally a simulation. As was highlighted in the introduction, our system is specifically designed to support simulations that consist of multiple heterogeneous games played by agents using a diverse set of strategies. For example, a simulation may consist of 100 rounds of the Minority Game followed by 100 rounds of the Iterated Prisoners Dilemma. Further, the participants in this game may be broken down such that 30\% are using a random strategy, 40\% are using a Tit-for-Tat strategy, and 30\% are using a random strategy. Interwoven into this model, we also envisage provision being made for periodic adaption/evolution steps whereby the participants are able to adjust/change their current strategy based on diverse means (ranging from random selection to a performance review or even the use of some form of genetic programming model).

Specifically, we envisage games to vary on the following axes:
\begin{itemize}
\item \textbf{Number of Players}: Two-Player, Multi-Player
\item \textbf{Moves}: Simultaneous, Sequential
\item \textbf{Payoff}: Zero-Sum, Non-Zero-sum, Rankings, Range of values (bounded and unbounded) 
\item \textbf{Identity}: Known, Unknown, Irrelevant
\item \textbf{Communication Between Players}: Possible, Not possible
\item \textbf{Topology}: Spatial, Non-spatial
\end{itemize}

Further, we envisage tournaments to vary on the following axes:
\begin{enumerate}
\item \textbf{Communication Between Players}: Possible, Not possible
\item \textbf{Game Order}: Ordered (known to players), Ordered (unknown to players), Random 
\item \textbf{Strategies}: Fixed-Bag-Fixed-Choice, Fixed-Bag-Random-Choice, Evolutionary Adaptation, Machine-learning adaptation
\end{enumerate}

Note: These are not the only possible axes or options on the axes. Rather, these are the options that we are currently implementing. To represent these axes in a simulation, we have fixed on the following model. A simulation is a \emph{tournament} that is played between a set of \emph{agent} players who employ a fixed range of \emph{strategies} (this can include meta-strategies that combine multiple sub-strategies). Each tournament consists of a number of \emph{games} to be played, each of which consists of one or more \emph{rounds}. Each round of a game involves an agent using their strategy to make a \emph{move}. Each move is a bid / choice made by the agent that is relevant to the game being played. From here on, we refer to agents and players synonymously unless specifically required for reasons of clarity.

\subsection{Decomposing Games}
\label{decomposing}

\begin{figure}
    \includegraphics[scale=0.27]{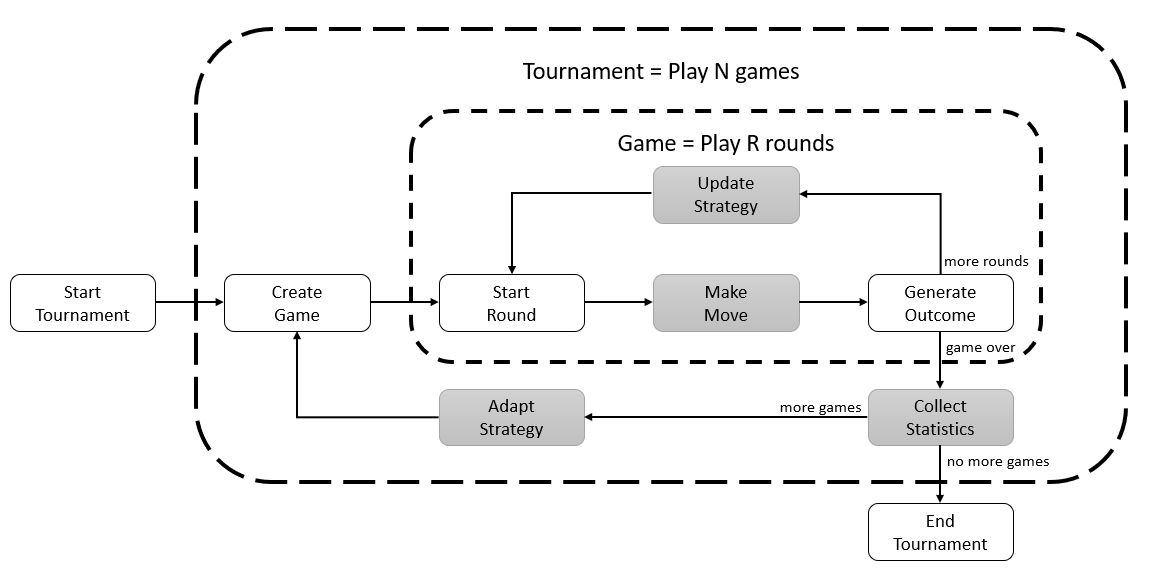}
    \caption{Stages of a Game}\label{fig:stages}
\end{figure}

Based on our abstract model of a simulation, we propose, in figure \ref{fig:stages}, a generalized workflow for the execution of a tournament:
\begin{itemize}
\item \emph{Start Tournament}: The tournament specification (what games, how many rounds, what agents, and what strategies) is loaded and the simulator initialized.
\item \emph{Create Game}: A game is created based on the corresponding specification and agents are linked to it.
\item \emph{Start Round}: Triggers the start of a round. This may involve transmission of a state to each agent or simply a request to make a move.
\item \emph{Make Move}: Here the agent must decide on its move which is based on the associated strategy. The agents move is submitted to the game.
\item \emph{Generate Outcome}: The game, once all required moves are made or a timer expires, decides on the outcome of the round (who wins and who loses, what is the payoff) and communicates this to the agents.
\item \emph{Update Strategy}: Here, the agent has the opportunity to update its strategy based on the outcome.
\item \emph{Collect Statistics}: When there are no more rounds in a game, the simulator gathers all specified statistics into a single resource that is stored for later analysis.
\item \emph{Adapt Strategy}: If there are more games to be played, then there is a chance for the agents to adapt their strategies. This may make use of some of the statistics gathered in the previous step. Adaptation can take any form that is appropriate, from parameter tuning to the use of evolutionary techniques, to machine learning techniques.
\item \emph{End Tournament}: The tournament finishes, statistics are collated into a data set that is released as a set of files and any requested summary statistics are presented to the user.
\end{itemize}
Of the above list of game stages, we believe that some are the responsibility of the simulator, some the game, and some by the strategy employed by the agent playing the game. For example, we feel that the stages in figure \ref{fig:stages} that are shaded are the responsibility of the agent / strategy. While the others are the responsibility of the game / simulator. The delineation allows us to cleanly separate game design from strategy design which in turn allows us to develop generalized strategies that can be used to play a diverse range of games.

\subsection{Generalized Game Design}
\label{sec:game}
To facilitate the integration of heterogeneous game types, we have attempted to create an abstract model of a game that can be customized as necessary. 

\begin{figure}[ht]
	\includegraphics[scale=0.55]{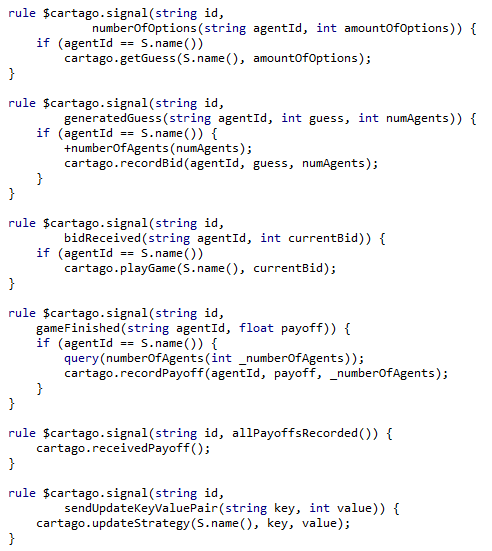}
    \captionof{figure}{Interaction of Players with Cartago Artifacts}
    \label{snippet:player-interaction-with-cartago}
\end{figure}

Underpinning the model is the view of the player as an agent that interacts with artifacts within the game environment. These artifacts serve as enablers / constrainers of agent moves. Through the artifacts, each game specifies whether agents can recognize each other, whether there is an ordering to moves, how many rounds exist in a game, whether each round has an entry and exit condition, whether payoff / penalty is dealt to the agents after every round or it accumulates through the game. Thus, each game is envisioned as a configuration of artifacts. This allows new games to be created without having to re-write many common aspects of game playing. Figure~\ref{snippet:player-interaction-with-cartago} shows a sample code that is used to implement \textit{both}, Iterated Prisoner's Dilemma as well as the Minority Game. This reconfiguration focused approach is extended deeper into agent-strategies as well, as shall be seen in the next sub-section. The artifact-based reconfiguration allows for another feature, evolutionary spread of features across agents. This means that agents that perform well can have their strategies copied and modified by other agents (via mechanisms such as clonal plasticity~\cite{Nallur2018}), thus leading to evolutionary pressure on strategies.

\subsection{Generalized Strategy Design}
\label{sec:strategy}
In order that strategies can be re-used across games, as well as new strategies added to a game, all strategies conform to an interface that agents can use to generate moves.

\begin{center}
	\includegraphics[scale=0.45]{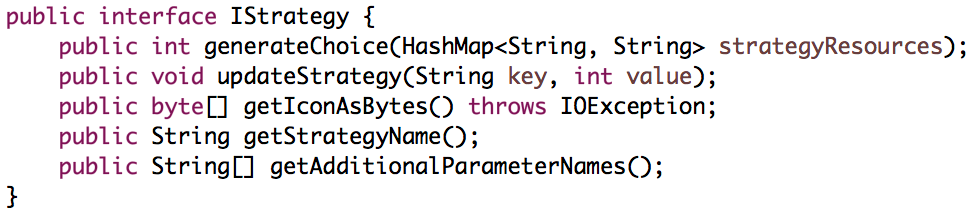}
    \captionof{figure}{Snippet of Strategy Interface}
    \label{snippet:strategy-interface}
\end{center}
The \textit{generateChoice} and \textit{updateStrategy} methods are self-explanatory. The \textit{strategyResources} map passed to \textit{generateChoice} contains external data which may be required by the agent in decision-making. The parameters passed to \textit{updateStrategy} are simply to be stored within the strategy object's internal resources map as a key-value pair. The remaining methods are required by the GUI for ensuring it reacts dynamically to user selection of strategies.

\subsection{Worked out Example}
The Iterated Prisoner's Dilemma (IPD) and the Minority Game (MG) are now compared to provide a concrete view of the generalization via artifacts.
\begin{table}[]
\centering
\begin{tabular}{@{}lll@{}}
\toprule
                  & \textbf{MG}             & \textbf{IPD}  \\ \midrule
Number of players & 3 or more (always odd)   & 2            \\
Moves             & Simultaneous             & Simultaneous \\
Amount of Payoff  & Fixed					 & Fixed        \\
Identity          & Irrelevant               & Known        \\
Comm b/w players  & Possible                 & Not Possible \\
Topology          & Non-spatial              & Non-spatial \\ \bottomrule
\end{tabular}
\caption{Comparing Minority Game and IPD}
\label{tbl:compare-ipd-mg}
\end{table}
IPD and MG share three artifacts (see Table~\ref{tbl:compare-ipd-mg}), while differing on three. Their implementation, therefore, is a simple composition of parametrized artifacts, which leads to faster and repeatable game creation.

\begin{figure}
	\includegraphics[scale=0.5]{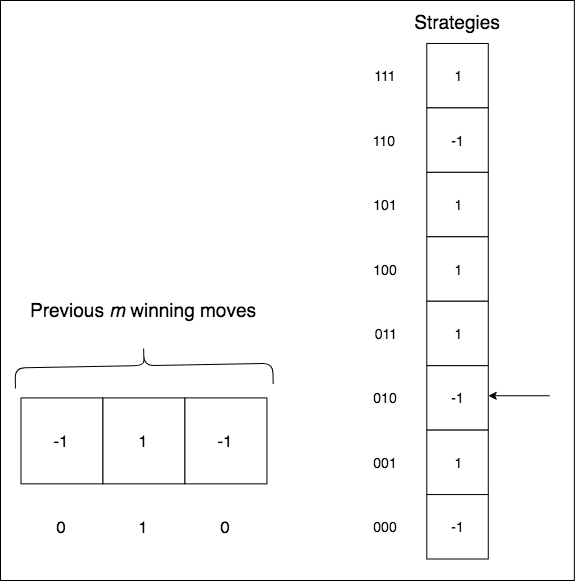}
    \caption{BestPlay example for m=3}
    \label{fig:bestplay}
\end{figure}

\paragraph{Strategy Generalization:}
Generalizing strategies such that they can be re-used across games is a little more involved. Quite often, the strategy is very closely tied to the rules of the game or the number of available options to the agents. 
BestPlay~\cite{Challet1997} is a strategy created by the creators of the Minority Game. Recall that the MinorityGame (MG) consists of a population of \texttt{N} (odd) individuals, who have to make a simultaneous binary choice. The group that is in the minority, after making a choice, is the winner that receives a payoff. Although simple in its setup and play, the dynamics in MG has been used in multiple fields such as econophysics~\cite{Bianconi2009,Zapart2009}, multi-agent resource allocation~\cite{Metzler2003,Li2004}, emergence of cooperation~\cite{Dhar2011}, and heterogeneity~\cite{Greenwood2009,Mello2008}. The BestPlay strategy utilizes two pieces of information:\textbf{(a)} Memory of the winning moves from previous \texttt{m} rounds; \textbf{(b)} Vector containing a pool of strategies. The vector, for each player, is of length $2^{m}$ and is used to decide what to play next. Since MG allows only two moves, ${1,-1}$, the previous \texttt{m} winning moves can be encoded as a binary number (see Figure~\ref{fig:bestplay})
The binary number can then be converted to an integer corresponding to a position in the strategy vector. In Figure~\ref{fig:bestplay}, the strategy vector yields $-1$, as $010$ corresponds to the integer $2$, giving the result in the third position of the vector. Due to the nature of the encoding strategies, BestPlay is only applicable to games that involve selecting one out of two choices in every move. Since our intention is to allow the same strategy to be used across as many games as possible, we created a generalized implementation of BestPlay that retains its fundamental structure, but is not limited by two choices.

\paragraph{Generalizing BestPlay:} The strategy vector can be generalized using a simple theorem that demonstrates how an \texttt{n}-ary code may be converted into an integer~\cite{Moll2012}. Given $a,b \in \mathbb{N}$ with $b \textgreater 1$, there exist non-negative integers $x_{0},x_{1}...x_{n}$ such that 
\[
a = x_{0} + x_{1}b + x_{2}b^{2} + \ldots + x_{n}b^{n} \]
with
$ 0 \leq x_{i} < b $ and $ x_{n} \neq 0 $

\begin{figure}
	\centering
	\includegraphics[scale=0.75]{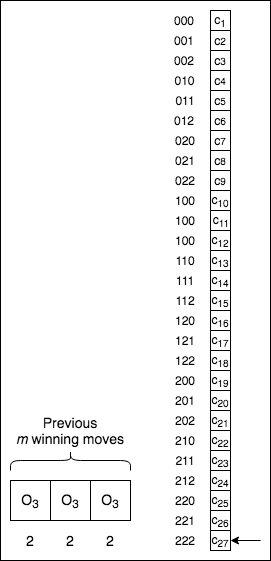}
    \caption{BestPlay example for m=3 and q=3}
    \label{fig:best-play-three}
\end{figure}

Consider a game which presents three options $(O_{1},O_{2},O_{3})$ to a player. The strategy vector would now be $q^{m}$ as opposed to $2^{m}$, where $q$ corresponds to the options available to a player. Continuing with the previous example of $m=3$, the strategy vector now has a length of $3^{3} = 27$. The choices $(O_{1},O_{2},O_{3})$ will be represented as options ${0,1,2}$. If the previous $m$ winning moves were ${O_{3},O_{3},O_{3}}$, which corresponds to code $222$, then BestPlay would index into the $26^{th}$ position of the strategy vector (see Figure~\ref{fig:best-play-three}).

This results in an implementation of the BestPlay strategy, as shown in Figure~\ref{snippet:best-play-constructor} and Figure~\ref{snippet:best-play-move}, playable in games with arbitrary number of choices.
\begin{center}
	\includegraphics[scale=0.30]{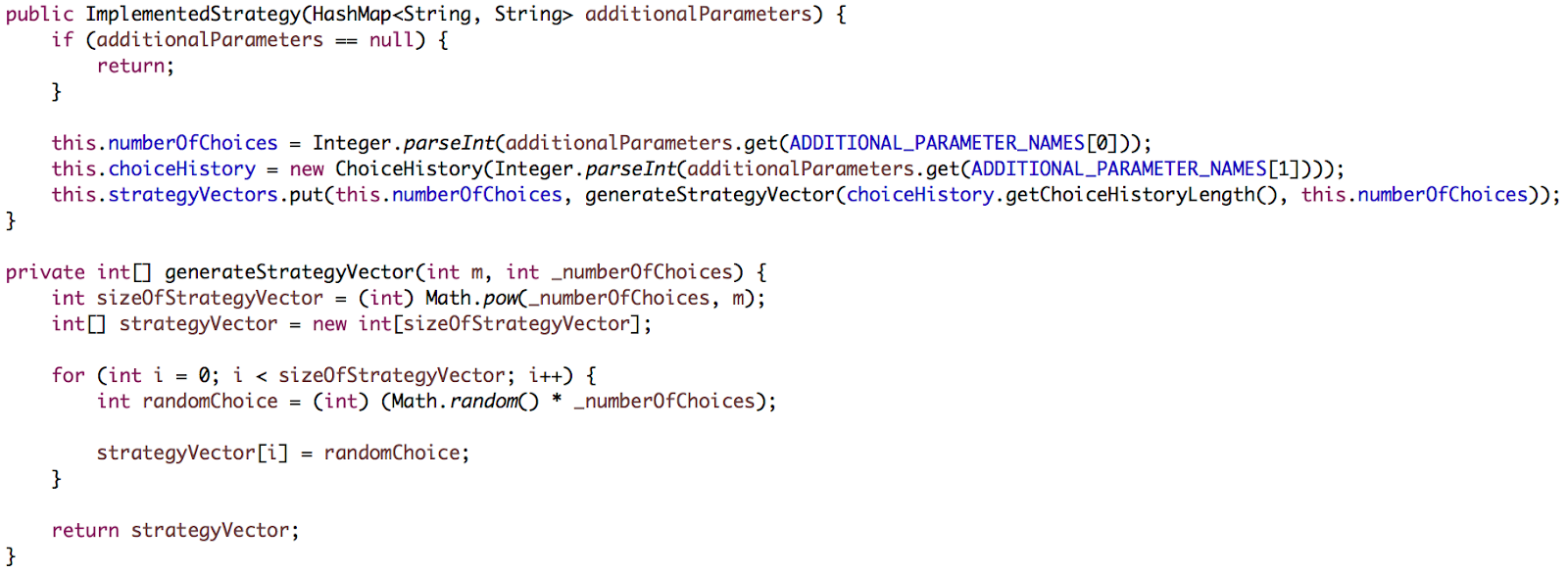}
    \captionof{figure}{Constructor for BestPlay Strategy}
    \label{snippet:best-play-constructor}
\end{center}
\begin{center}
	\includegraphics[scale=0.35]{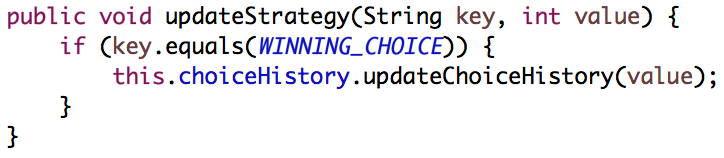}
    \captionof{figure}{Generating a move using BestPlay Strategy}
    \label{snippet:best-play-move}
\end{center}

\section{\uppercase{An AOP-based Game Simulator}}
\label{sec:framework}
The goal of this work is to develop a flexible generalized game simulator through which we can study various properties of games and associated strategies. Specifically, we want to enable the development of a simulator that can support a heterogeneous set of games being played using a diverse suite of generalized strategies. We also wish to explore how these generalized strategies perform when applied to multiple game types that are played sequentially.

Typically, such simulators are implemented using languages such as NetLogo or using some form of generalized programming language (e.g. Java, C). In contrast, our approach is to explore the use of Agent-Oriented Programming (AOP) languages \cite{Shoham1993} and related technologies.  Specifically we will use the ASTRA language \cite{Collier2015}, which is a variation of AgentSpeak(L) \cite{Rao1996} together with CArtAgO \cite{Ricci2006} a framework that supports the modeling of the agents environment in terms of shared objects known as artifacts. This allows for modelling phenomena such as cultural learning where agents can copy strategies from successful neighbours.

\begin{figure}
	\includegraphics[width=\columnwidth]{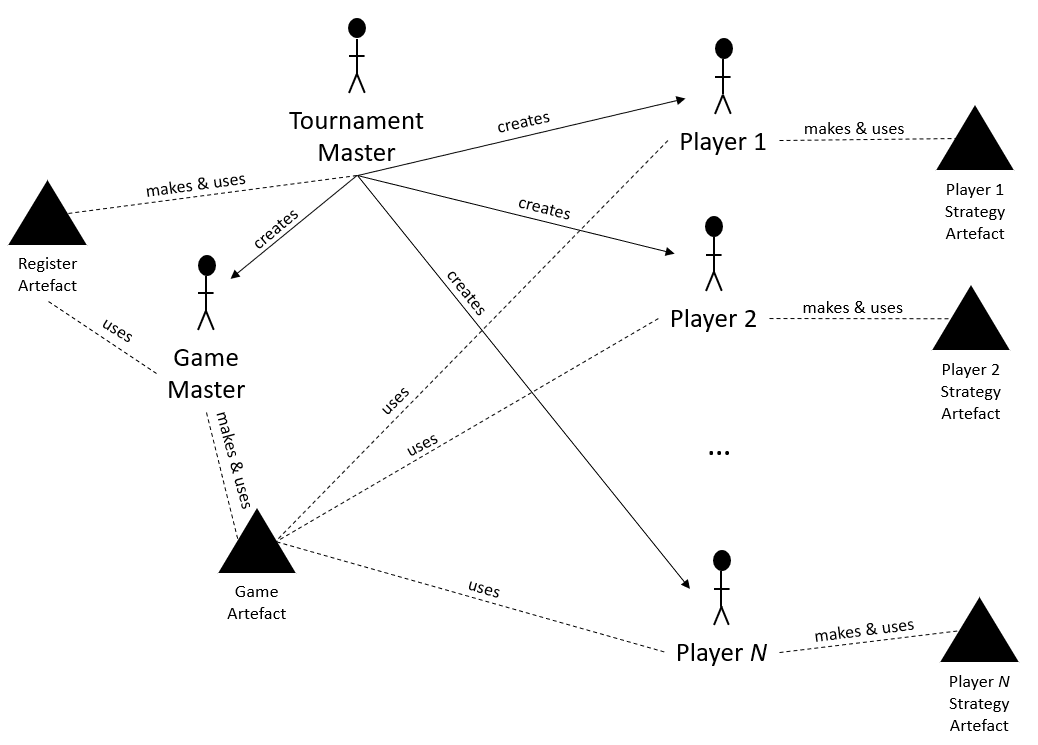}
    \caption{AOP-based Deployment}\label{fig:deploy}
\end{figure}

A high level view of the proposed framework is highlighted in figure \ref{fig:deploy}. In this figure, the stick people are agents, and the triangles are (CArtAgO) artifacts. In line with section 2, the game artifact specifies a generalized interface through which agents can interact with a game (e.g. make move) and the strategy artifacts are generalized artifacts that can be used to instantiate and use specific game strategies (e.g. random play, best play, ...). The register artifact provides a centralized list of players and their availability. When started, the \emph{Tournament Master} reads the tournament specification and creates an initial community of \emph{Player} agents (who create associated strategy artifacts). These players are added to the registry. The \emph{Tournament Master} then creates a \emph{Game Master} who sets up a Game artifact that the \emph{Player} agents connect to. The \emph{Game Master} is responsible for the  execution of the game. Its first task is to select a set of players to play the game. This is dependent on the selection policy adopted, and can be either a random selection of players of specific types or a fixed set of players. These players are invited to join the game, the game is played, and at the end, the \emph{Game Master} informs the players that the game is over (causing the players to disconnect from the associated game artifact).

Our initial plan is to generate generic implementations of the \emph{Tournament Master}, \emph{Game Master}, and \emph{Player} agents. These implementations will allow enough configuration to support their use with any game / basic game strategy. However, the motivation for using an AOP language to implement this is that, in the future (see our roadmap \ref{sec:roadmap}) we intend to start changing the \emph{Player} agent to support more complex game playing behaviours. Due to the clear separation of concerns enforced through the use of an AOP language, we believe that this will be easier to achieve than if we had used a standard general purpose programming language. The current implementation can be accessed at:~\texttt{https://gitlab.com/aop-arena} 

\section{\uppercase{Roadmap}}
\label{sec:roadmap}
This paper describes a prototype, and is therefore only able to describe features that we have already implemented. However, we have a roadmap of features that we are working on implementing. Currently, all implementation is in the form of ASTRA agents and CArtAgO artifacts. However, this requires re-compilation of code whenever any aspect of the experiment changes. For easier experiment design and setup, we expect that more aspects need to be made into pure configuration text, so that non-coding-specialists can also use Arena. The features we expect to add to Arena are:

\begin{enumerate}

\item \textbf{Open System with agents that enter and exit a tournament}: Complex domains such as urban modelling often require that simulations be `open', \textit{i.e.,} agents must be able to enter, exit and re-enter a simulation.

\item \textbf{Game Description Language}: A human-readable and machine-parseable language for describing game rules, in terms of setup, number of players, sequential or simultaneous, competitive or cooperative or coalitional, kinds of payoffs, number of rounds, time-bound or not etc.

\item \textbf{Player Description Language}: Players can diverge on the kinds of strategies that they implement, \textit{i.e.,} they may be adaptive agents that adapt their strategies or machine learning agents that have \textit{one} strategy that continuously adapts itself, or agents that possess a bag-of-strategies which they play in some order. 

\item \textbf{Tournament Description Language}: A human-readable and machine-parseable language for describing the order of games, number of repetitions for each game, open or closed system (can new agents enter in the middle of a tournament?)

\item \textbf{Detection of Emergence within Tournament}: The presence of features such as known identity of players, multiple heterogeneous games, sophisticated strategies such as evolution / machine learning could lead to the emergence of `agreements', unofficial rules, etc. which would be valuable to detect. We aim to incorporate tournament-wide emergence detection~\cite{OToole2017} mechanisms to allow for automated monitoring of large-scale tournaments.

\end{enumerate}


\section{\uppercase{Conclusions}}
\label{sec:conclusion}
This paper reports on a first prototype of a generalized game playing framework called \textit{Arena}, that allows for the same agents to play multiple heterogeneous games. Arena currently allows strategies, game rules, and players to evolve independently of each other. As of this writing, we have implemented the Iterated Prisoner's Dilemma, Minority Game, Linear Public Goods Game, with agents that can play multiple rounds of each game, based on a tournament configuration. Strategies such as Tit-for-Tat, BestPlay have also been implemented in a generalized manner such that they can be re-used across games, by merely querying game parameters. We expect a more mature version of the framework to be a valuable tool, both for the agents community as well as game theory researchers.


\bibliographystyle{ACM-Reference-Format}  
\bibliography{game-theory-agents}  

\end{document}